\documentclass[twocolumn,fleqn,10pt]{wlscirep}
\usepackage[utf8]{inputenc}
\usepackage[T1]{fontenc}
\usepackage{color}
\usepackage{relsize}
\title{Direct observation of zero modes in a non-Hermitian nanocavity array}

\author[1]{Flore Hentinger}
\author[1]{Melissa Hedir}
\author[1]{Bruno Garbin}
\author[2]{Mathias Marconi}
\author[3]{Li Ge}
\author[1,4]{Fabrice Raineri}
\author[1]{Ariel Levenson}
\author[1,*]{Alejandro M. Yacomotti}
\affil[1]{Centre de Nanosciences et de Nanotechnologies, CNRS, Universit\'e Paris-Sud, Universit\'e Paris-Saclay, 10 Boulevard Thomas Gobert, 91120 Palaiseau, France}
\affil[2]{Universit\'e C\^ote d'Azur, Institut de Physique de Nice, CNRS-UMR 7010, Sophia Antipolis, France}
\affil[3]{Department of Physics and Astronomy, College of Staten Island, CUNY, Staten Island, New York 10314, USA and Graduate Center, CUNY, New York, New York 10016, USA}
\affil[4]{Universit\'e Paris-Diderot, 75205 Paris Cedex 13, France}
\affil[*]{Corresponding author: alejandro.giacomotti@c2n.upsaclay.fr}

\begin{abstract}
Zero modes are symmetry protected ones whose energy eigenvalues have zero real parts. In Hermitian arrays, they arise as a consequence of the sublattice symmetry, implying that they are dark modes. In non-Hermitian systems, that naturally emerge in gain/loss optical cavities, 
particle-hole symmetry prevails instead; the resulting zero modes are no longer dark but feature $\pi/2$ phase jumps between adjacent cavities. Here we report on the direct observation of zero modes in a non-Hermitian three coupled photonic crystal nanocavity array containing quantum wells. Unlike the Hermitian counterparts, the non-Hermitian zero modes can only be observed for small sublattice detuning, and they can be identified through far-field imaging and spectral filtering of the photoluminescence at selected pump locations. We explain the zero mode coalescence as a parity-time phase transition for small coupling. These zero modes are robust against coupling disorder, and can be used for laser mode engineering and photonic computing.

\end{abstract}
\begin{document}

\flushbottom
\maketitle
\thispagestyle{empty}

%\noindent Please note: Abbreviations should be introduced at the first mention in the main text – no abbreviations lists. Suggested structure of main text (not enforced) is provided below.

%\section*{Introduction}
%
%The Introduction section, of referenced text\cite{Figueredo:2009dg} expands on the background of the work (some overlap with the Abstract is acceptable). The introduction should not include subheadings.
%

\section*{Introduction}

Majorana zero modes, a captivating concept originally proposed in the study of neutrinos, have intrigued physicists over the past eighty years. Being their own anti-particles and hosting non-Abelian braiding properties, their experimental demonstration is being actively pursued in high-energy physics and condensed matter physics \cite{hasan_colloquium:_2010,qi_topological_2011,alicea_new_2012,beenakker_random-matrix_2015,nayak_non-abelian_2008}. The existence of these zero-energy excitations is warranted by particle-hole symmetry, in the form that the (Hermitian) Hamiltonian anti-commutes with an anti-linear operator \cite{rivero_chiral_2021}.

Recently, there have been several proposals to realize particle-hole symmetry in non-Hermitian systems \cite{feng_non-hermitian_2017}, especially on integrated photonic platforms where the spatial arrangement of optical gain and loss
\cite{PhysRevA.95.023812}, 
as well as asymmetric couplings between different elements \cite{ malzard_topologically_2015}, plays an important role. These findings are quite surprising because photons are bosons and cannot form \lq\lq particle-hole'' pairs in general. However, by realizing that these effective \lq\lq particles" and \lq\lq holes'' have complex energies in a non-Hermitian system, they do not need to adhere to Fermi-Dirac statistics as their condensed matter counterparts do. To highlight this difference, we will refer to such symmetries as non-Hermitian particle-hole (NHPH) symmetry.

Although the resulting zero modes of NHPH symmetry differ from Majorana zero modes in several key aspects, they have two  desirable properties in many photonic applications: their symmetry protection not only exists at the origin of the complex energy plane but also extends to the entire imaginary axis; they can also be conveniently excited in standard arrays of optical cavities or waveguides, without requiring the existence of Hermitian counterparts when non-Hermiticity is removed.

Thanks to the flexibility of designing optical elements, these properties of photonic zero modes can also be induced by pseudo-anti-Hermiticity \cite{rivero_pseudochirality_2020}: $
\eta H^\dagger \eta^{-1} = -H$, where $H$ is the Hamiltonian and $\eta$ a linear operator. 
If we consider a system consisting of two sublattices $A$ and $B$, where couplings only take place between two cavities belonging to different sublattices,  
pseudo-anti-Hermiticity coincides with NHPH symmetry 
%(i.e., $\{\eta K,H\}=0$, where $K$ is the complex conjugation and $\eta=P_A-P_ B$, $P_{A,B}$ being the projection operators onto the two sublattices)  
when $H$ is symmetric, but it is distinct otherwise such as in a topological insulator laser \cite{bandres_topological_2018,zhao_non-hermitian_2019}, where an effective gauge field is realized by staggered couplings in a two-dimensional array.

%Thanks to the flexibility of designing optical elements, these properties of photonic zero modes can also be induced by pseudo-anti-Hermiticity \cite{rivero_pseudochirality_2020}: $
%\eta H^\dagger \eta^{-1} = -H$, where $H$ is the Hamiltonian and $\eta$ a linear operator. 
%If we consider a system consisting of two sublattices $A$ and $B$, where couplings only take place between two cavities belonging to different sublattices,  
%pseudo-anti-Hermiticity coincides with NHPH symmetry (i.e., $\{\eta K,H\}=0$, where $K$ is the complex conjugation and $\eta=P_A-P_ B$, $P_{A,B}$ being the projection operators onto the two sublattices)  when $H$ is symmetric, but it is distinct otherwise such as in a topological insulator laser \cite{bandres_topological_2018,zhao_non-hermitian_2019}, where an effective gauge field is realized by staggered couplings in a two-dimensional array.

Despite these theoretical advances in the non-Hermitian domain, the observations of photonic zero modes with the aforementioned properties have been restricted to arrays that resemble their Hermitian counterparts, such as the Su-Schrieffer-Heeger (SSH) lattice \cite{poli_selective_2015,zhao_topological_2018,song_breakup_2019, Pan_2018}. A recent demonstration of a laser mode switching in a coupled photonic crystal cavity could also be explained on the basis of Hermitian dark modes \cite{doi:10.1063/5.0006767}. Active photonic crystal cavity arrays are outstanding platforms to access the non Hermitian realm because they naturally enable in-situ realization of gain/loss configurations and coupling engineering \cite{haddadi2014}. Remarkably, they have recently led to the observation of exceptional points in two coupled nanocavities \cite{Kim:2016aa,Takata:21}. 

In order to show the potential of exciting and controlling a photonic zero mode in a broader range of systems, especially those without a topological origin \cite{qi_defect_2018,PhysRevA.95.023812}, here we report on its observation in a minimal system consisting of three coupled photonic crystal cavities with NHPH symmetry.  Interestingly, while Hermitian zero modes are robust against frequency detuning between the two extreme cavities (sublattice $A$) and the central one (sublattice $B$) in a linear array --that we refer to as sublattice detuning, $\Delta\omega$--, non Hermitian zero modes are not. In our coupled cavity system, the intercavity coupling $g$ is modified by design, allowing to feature both large ($|g|\gtrsim |\Delta\omega|$) and small  ($|g|< |\Delta\omega|$) coupling regimes. Hereby we will show that, when entering into the large detuning phase, the zero mode first looses its properties because it is no longer NHPH symmetry protected, and eventually coalesces with another lattice mode through a parity-time phase transition. 

This paper is organized as follows. In Section \ref{S1} we provide a simple theoretical framework based on coupled mode theory (CMT) to understand zero modes in gain/loss cavity arrays warranted by NHPH symmetry. In Section \ref{S2} we describe our photonic crystal  three cavity array with controllable coupling by means of the so-called barrier engineering technique. We also provide an experimental characterization of the linear Hermitian modes though resonant scattering experiments. In Section \ref{S3} we move into non Hermitian mode characterization by incoherently pumping the system, and we report on the direct observation of the zero mode in the small sublattice detuning regime. Such observation is based on photoluminescence intensity maps under the spatial scanning of the pump spot combined with a Fourier imaging technique. In Section \ref{S4} we provide a detailed analysis of the phase transitions in the system as both the coupling and the sublattice detuning are changed; a systematic comparison with linear and nonlinear CMT modeling allows us to explain the coalescence of the zero mode as a result of a parity time spontaneous symmetry breaking mechanism. Conclusions are given in Section \ref{S5}.
%Some recent results to be mentioned: 
%
%- A. Fiore's laser mode switching using dark modes in a 3-coupled cavity system \cite{doi:10.1063/5.0006767}
%
%-Notomi's Observation of PT exceptional point in coupled PhC cavities \cite{Takata:21}

\section{Theoretical background: non-Hermitian zero modes}
\label{S1}

\begin{figure*}[!h]
\centering
\includegraphics[trim=0.5cm 3cm 0cm 1cm,clip=true,scale=0.6,angle=0,origin=c]{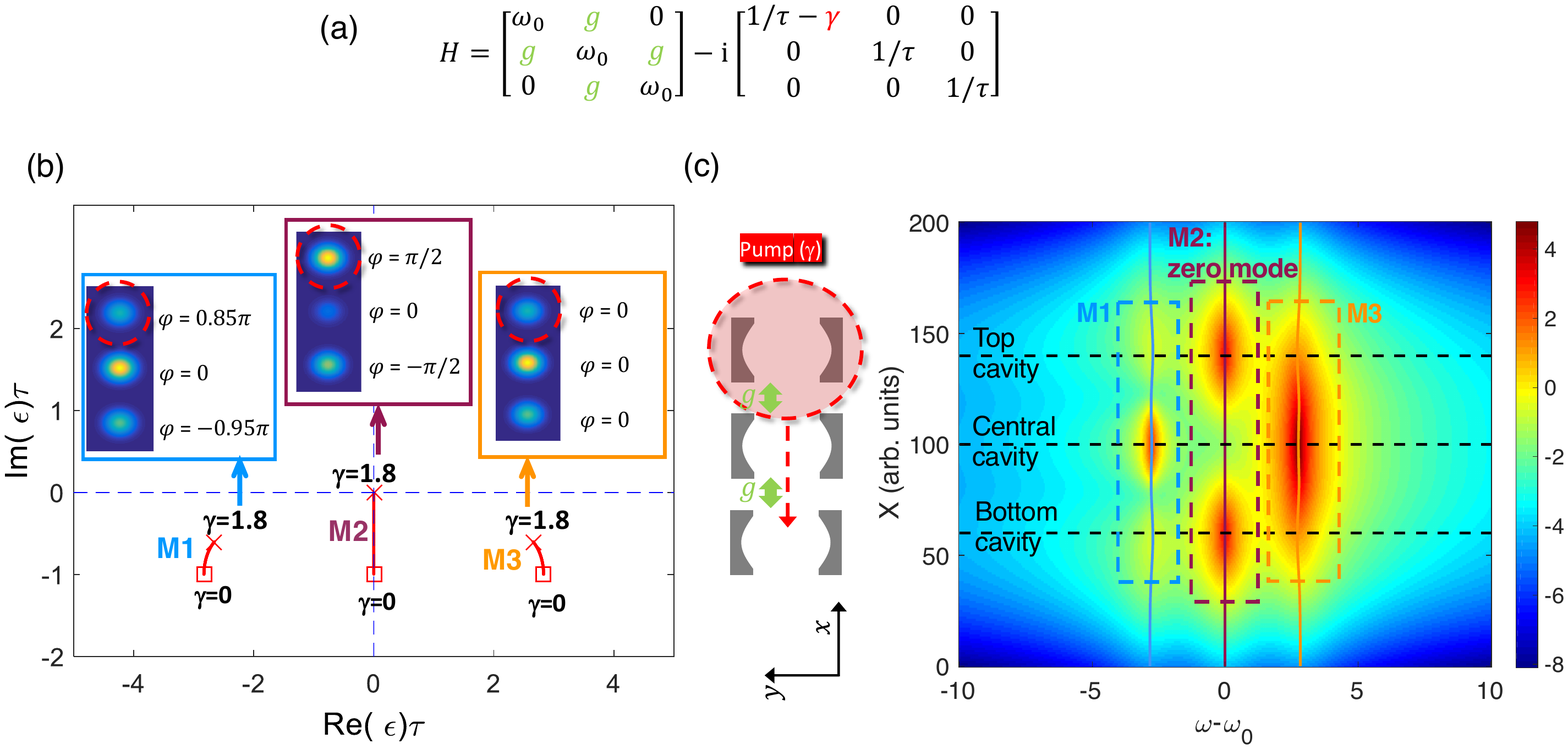}
\caption{Non Hermitian zero mode in a three coupled cavity array: CMT model. (a) Non-Hermitian Hamiltonian. (b) Eigenvalues and eigenvectors. Inset: intensity and phase spatial distribution at $M2$-laser threshold. (c) Logarithmic spectral intensity as a function of the center of a gaussian pump spot, computed from Eq. \ref{eq:I(w)} in the text. The pump spot profile is $P(x ;X)=\exp[-(x-X)^2/\sigma^2]\gamma$, with $\sigma=30$ and $\gamma=1.5$, i.e. below laser threshold. Solid lines correspond to $\operatorname{Re}(\epsilon_j)$ for $j=1$ ($M1$, blue), $j=2$ ($M2$, burgundy) and $j=3$ ($M3$, orange).} \label{fig1}
\end{figure*}

A simple theoretical modal analysis of evanescently coupled cavity lattices can be carried out in the framework of the Coupled Mode Theory (CMT) formalism. In the case of $N$ resonant optical cavity lattices, the CMT is valid under the hypothesis of negligible coupling between non-adjacent cavities and weak coupling overall. It assumes that the system can be accurately described with both the isolated (real) cavity frequencies $\omega_{n{\in [1,N]}}$ and the coupling strength to their neighboring cavities. The resulting hybrid mode frequencies and field distributions become the eigenvalues and eigenvectors of a Hamiltonian operator $H$, which does not need to be Hermitian. This is naturally the case of optical cavities in the presence of loss and/or gain \cite{feng_non-hermitian_2017}: a lattice can be described in CMT by a Hamiltonian $H$ whose matrix elements are $H_{nn}=\tilde\omega_n$ and $H_{nm}=g_{nm}$ ($n \neq m$), where $\tilde\omega_n$ are the complex frequencies of the isolated cavities and $g_{nm}$ are the nearest-neighbor inter-cavity coupling parameters. In a closed Hermitian system, $\tilde\omega_n=\omega_n$  are  real, while in  a non-Hermitian gain/loss optical system $\tilde\omega_n=\omega_n-i/\tau_n$, where $\tau_n$ is the $n-$th cavity lifetime, which can be negative for net gain.
Also $g_{nm}$ can become complex in general in a non-Hermitian framework \cite{https://doi.org/10.1002/lpor.201700113}; here we will consider real $g_{nm}$ for simplicity, {\em i.e.} we will neglect loss splitting \cite{Atlasov_2008,hamel2015}.  

Although our analysis can be extended to large cavity networks, in this work we will focus on a small array of three coupled cavities as the minimal system containing a zero mode. Figure \ref{fig1} illustrates a simple case where three cavities are aligned along the $x$-direction, while the resonant intracavity field oscillates back and forth in the $y$-direction (see schematics in Fig. \ref{fig1}(c)]. We assume they all have the same resonant frequency $\omega_0$ chosen as the reference frequency, nearest neighbor coupling $g_{nm}=g=2$, and intrinsic field decay time $\tau_0=1$ due to optical losses. Let us point out that the zero mode has the same frequency as a standalone cavity. One of the extreme cavities, say the top one, is incoherently pumped, therefore introducing a variable gain $\gamma$, see Fig. \ref{fig1}(a). The complex eigenvalues of $H$ and their evolution with increasing $\gamma$ are depicted in Fig. \ref{fig1}(b). Red symbols correspond to the eigenvalues $\varepsilon_j$ ($j=1,2,3$) evolving as the pump is increased from $\gamma=0$ (red square at the starting point) to $\gamma=1.8$ (red cross at the end point). We call $M1$ and  $M3$ the lowest and highest frequency modes, respectively, both having a nearly symmetric field distribution, while $M2$ is the central mode featuring $\pi/2$ phase jumps between adjacent cavities [Fig. \ref{fig1}(b), inset].

A zero mode that can be realized in photonics is often the result of the sublattice (or chiral) symmetry in Hermitian lattices, where eigenvalues are real. Hermitian chiral arrays ensure $\varepsilon_k=-\varepsilon_j$ and a zero mode with $j=k$ verifies $\varepsilon_j=0$. These are known as dark modes, because the intensity in one of the sublattices vanishes. On the other hand, eigenvalues are complex in general in non-Hermitian arrays. While chiral symmetry can still be realized in this case \cite{rivero_chiral_2021}, a more prevailing symmetry that leads to a zero mode in a non-Hermitian lattice is NHPH symmetry, giving rise to  $\varepsilon_k=-\varepsilon_j^*$. A zero mode ($j=k$), which always exists in the case of an odd number of cavities, leads to $\operatorname{Re}(\varepsilon_j)=0$. This is the case of $M2$, whose frequency does not depend on $\gamma$ because of the NHPH symmetry protection, while $M1$ and $M3$ frequencies do.  It is worthwhile noting that such a frequency change for $M1$ and $M3$ is a pure non Hermitian effect, not related to any nonlinear refractive index effect, as is usual in semiconductors; in Secs. \ref{S3} and \ref{S4} we will include such carrier-induced refractive index effects to model our experimental results more realistically. 

Unlike the Hermitian counterparts, the zero mode is no longer a dark one under single cavity-pumping, see Fig. \ref{fig1}(b): no light extinction occurs in the central one. The exact $\pi/2$ phase shift between adjacent cavities distinguishes the wave function of a zero mode from all other modes, and dramatically impacts the photoluminescence far-field imaging in experiments, as will be shown later on.

Non-Hermitian zero modes warranted by NHPH symmetry have the freedom to evolve along the $\operatorname{Im}(\varepsilon)$-axis. In particular, at $\gamma=1.8$,  $\operatorname{Im}(\varepsilon_2)$ also becomes zero, so that the gain compensates the losses. 
%For $\gamma\geq 1.8$, $M2$ enters into the laser regime, and the linear model would no longer be valid. 
Thus, the linear CMT model predicts zero-mode lasing in three coupled cavity systems as one of the extreme cavities is pumped \cite{PhysRevA.95.023812}. Even though the zero mode can be clearly excited and identified by spatial localization of a single pump spot, semiconductor refractive index nonlinearities will generally prevent laser operation (see Section \ref{S3}).% ({\color{red} See Supplementary Material, Section...???}). 

\begin{figure*}[!h]
\centering
\includegraphics[trim=1cm 2cm 0cm 1cm,clip=true,scale=0.6,angle=0,origin=c]{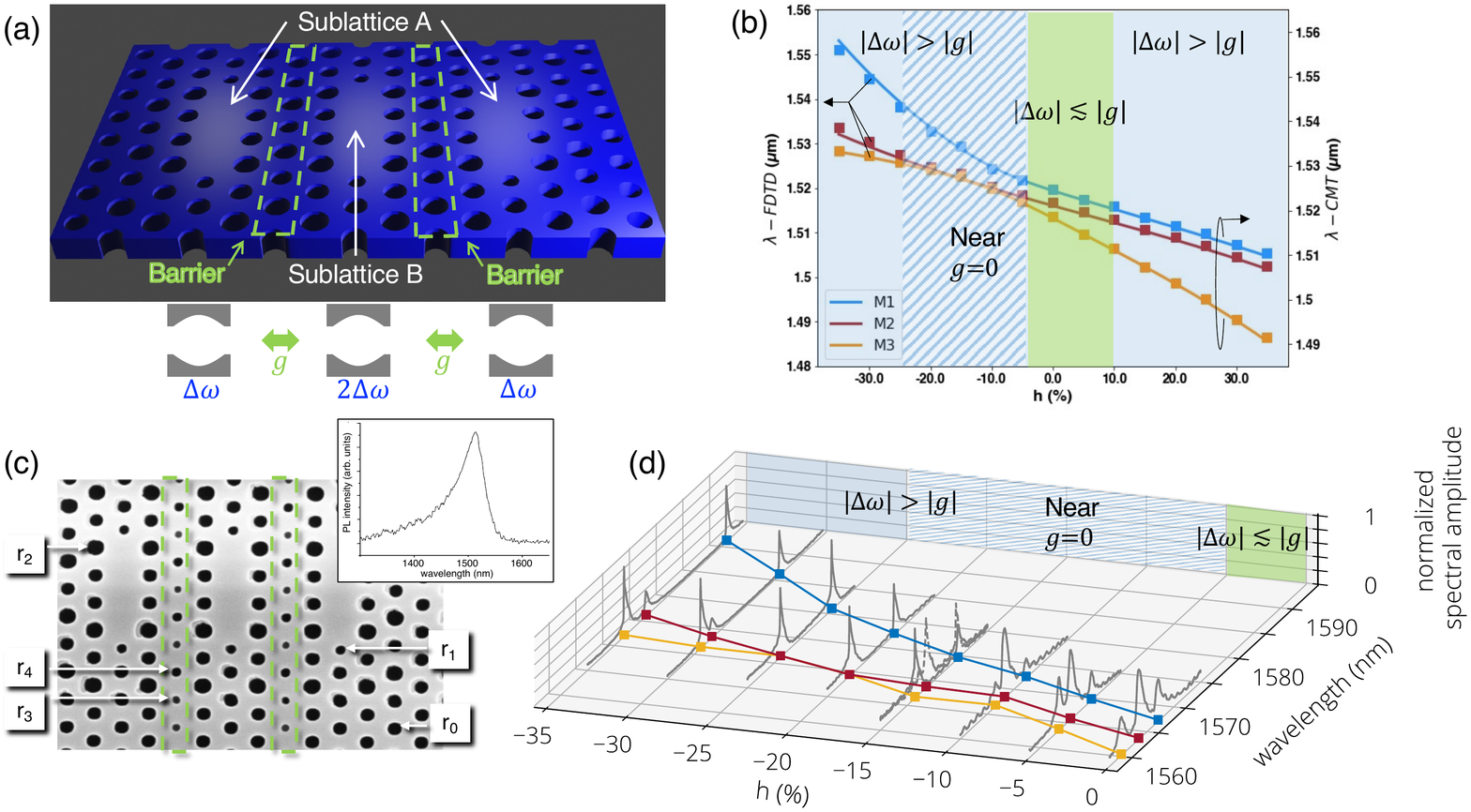}
\caption{Three coupled photonic crystal cavities. (a) Artist view of the system, featuring controllable coupling by means of the two barriers (highlighted with dashed boxes) in which holes are modified. Two sublattices A and B can be defined, where couplings only take place between cavities belonging to different sublattices. Bottom: schematic representation showing how the presence of the barriers modifies the cavity detuning. The sublattice detuning is then $\Delta\omega$. (b) 3D-FDTD simulation results of three coupled PhC cavities showing the evolution of the mode resonant wavelengths (symbols, left axis) as a function of the barrier parameter. Solid lines (right axis) are CMT predictions using polynomial approximations of $g(h)$ and $\Delta(h)$ obtained from two-coupled cavity FDTD simulations (see Section III, Supplementary Material). Two regions can be distinguished: the low (light green, $|g|\gtrsim |\Delta \omega|$) and high sublattice detuning regions (light blue, $|g|< |\Delta \omega|$); the near zero coupling region is highlighted with striped background. (c) SEM image of a sample of three coupled L3 InP-based photonic crystal cavities. Inset: QW photoluminescence. (d) Normalized reflectivity spectra as a function of the barrier parameter, from resonant scattering experiments. These results can be interpreted as the optical response of the coupled cavity system in the Hermitian limit. }  
\label{fig2}
\end{figure*}

Here we will restrict our analysis to the spontaneous emission regime --i.e. below laser threshold and neglecting amplification--, in which a linear non-Hermitian CMT is valid. We assume a (gaussian) pump spot profile $P(x ;X)$ centered at a given $X$-position, therefore $\gamma_n(X)=P(x_n; X)$, where $x_n$ are the central positions of the cavities. As a result, $\varepsilon_j$ depends not only on the pump power but also on the pump position, $\varepsilon_j=\varepsilon_j(X)$. The spontaneous emission in the cavities is $|f\rangle$  ($f_n\propto \gamma_n^{1/2}$), and the modal excitation amplitudes $f_j(X)=\langle \Phi_j| f\rangle$, $\langle \Phi_j|$ being the left eigenvectors of $H$. In this spontaneous emission regime the total emitted spectral intensity can be calculated as the incoherent superposition of the $N$ mode intensities, each one contributing with a Lorentzian peak of amplitude $f_j(X)$, resonant frequency  $ \operatorname{Re}[\varepsilon_j(X)]$ and width $ \operatorname{Im}[\varepsilon_j(X)]$: 
\begin{eqnarray}
I(\omega;X)&=&\left| \mathlarger{\mathlarger{\sum}}_j \frac{f_j(X)}{\left( \omega-\operatorname{Re}\left[ \varepsilon_j(X) \right] \right)i +  \operatorname{Im}\left[ \varepsilon_j(X) \right] } \right|^2 \\
 &\simeq&
\mathlarger{\mathlarger{\sum}}_j \frac{|f_j(X)|^2}{\left( \omega-\operatorname{Re}\left[ \varepsilon_j(X) \right] \right)^2 + \left( \operatorname{Im}\left[ \varepsilon_j(X) \right] \right)^2}.
\label{eq:I(w)}
\end{eqnarray}
%\cc{[I'm a little confused here. The expression here seems to suggest that the intensity of ASE increase linearly with the pump power, but shouldn't this dependence be exponential?]} {\color{blue}: you are right in that ASE implies superlinear increase of I as a function of the pump. Actually here we're neglecting stimulated emission for simplicity, then we approximate I as a linear function of carrier density, therefore as a linear function of pump. I removed "amplified" before, otherwise it's misleading, you're right. I also clarified: "Here we will restrict our analysis to the spontaneous emission regime below laser threshold"]}
%In what follows we will use the additional simplification that $|f_j|^2$ are approximately equal, which physically means that the PL curve is flat in the wavelength range of interest, and there is no spatial coherence in the PL emission. 
%In this way, the coupled cavity array behaves as a spectral filter of (amplified) spontaneous emission.
Figure \ref{fig1}(c) shows the spectral intensity map $I(\omega;X)$ under spatial scanning of a gaussian pump spot, computed from Eq. \ref{eq:I(w)}. The signatures of the zero mode are the two central lobes corresponding to $M2$ in Fig. \ref{fig1}(c). Such intensity map, together with systematic far-field measurements, constitutes a tool to experimentally investigate zero-mode radiation in the active cavity array.

\section{Three-coupled photonic crystal cavities: design and characterization of Hermitian modes}
\label{S2}

Photonic crystal (PhC) cavities with embedded quantum wells (QWs) are a suitable platform to experimentally investigate zero-mode photonics. This is due to the multiple degrees of freedom provided by the design parameters, as well as the intrinsic and controllable gain/absoption features. Three coupled PhC L3 cavities [three missing holes in the $\Gamma$K direction of a triangular air hole lattice, see Fig. \ref{fig2}(a)] are separated by three rows of holes in the $\Gamma $M direction, leading to evanescent coupling. The two extreme cavities of the linear array couple to the middle cavity only. In order to control the inter-cavity coupling we implement the so-called barrier engineering technique, by virtue of which the coupling strength (and even its sign) can be changed modifying the middle row separating two adjacent cavities \cite{haddadi2014,hamel2015}. We have designed the central hole-row in the barriers with radius $r_3= r_0 (1+h)$, $r_0$ being the hole radius of the underlying PhC lattice. We call the parameter $h$ the barrier perturbation. 
%We consider the range of $h$ between $-40 \%$ to $40 \%$; samples with $h\leq 0$ have been realized, which suffices to largely tune the inter-cavity coupling strength $g$. 
Importantly, $h$ has a strong impact on the cavity frequencies due to its influence on the effective refractive index surrounding the cavities. Since the barrier induces a frequency detuning $\Delta \omega (h)$ in a contiguous nanocavity, a good approximation is to consider the two extreme cavities as having the same frequency $\omega_0+\Delta \omega (h)$ and $\omega_0+2 \Delta \omega (h)$ for the central cavity [see schematics in Fig. \ref{fig2}(a), bottom, and Section III of the Supplementary Material for further details]. 

In order to predict the influence of the barrier parameter in the coupled mode structure we have carried out 3D-Finite difference Time Domain (FDTD) simulations. First, $g(h)$ and $\Delta \omega(h)$ have been obtained by polynomial fitting datasets of a two coupled cavity system separated by a barrier with perturbation $h$ (Fig. S1, Supplementary Material). Importantly, two regions of $h$ can be distinguished: the large detuning region, $|\Delta \omega(h)| > |g(h)|$ for $h< -5\%$ or $h> 10\%$, and the small detuning region, $|\Delta \omega(h)| \lesssim |g(h)|$ for $-5\%\lesssim h \leq 10\% $. 
Subsequently, 3D-FDTD simulation with three cavities have been carried out. The mode frequencies as a function of $h$ are depicted in Fig. \ref{fig2}(b), together with the Hermitian CMT predictions using the fitted parameters $\Delta \omega(h)$ and $g(h)$, showing very good agreement. 
%By studying the frequencies of the two eigenmodes (symmetric and antisymmetric), we can deduce the coupling strength between the two neighbour cavities and the cavity frequencies of both nanolasers and their dependency on the barrier parameter [?] (see Fig. 3(c)). 
%For three coupled cavities, the frequencies of all three cavities are modified due to barrier perturbation; we assume that they are shifted proportionally to the number of barriers that affects them. Therefore, the detuning Δω(h) between the middle cavity (which sees two barriers) and the extreme cavities (which see only one barrier each) is equal to the previously calculated frequency shift. (see Fig. 3(c)). In addition, we assume that the coupling between two adjacent cavities in the triple-cavity lattice is the same as with a double- cavity lattice. 

We have fabricated the PhC cavity array of Fig. \ref{fig2}(a) in a suspended InP membrane of 280 nm-thickness containing four InGa$_{0.17}$As$_{0.76}$P quantum wells (QWs), featuring a photoluminescence peak at $\lambda\approx 1514$ nm [Fig. \ref{fig2}(c)]. The details on the fabrication can be found in Ref. \cite{hamel2015}.  The hole radius is $r_0=0.266a$ and $a$, the period of the triangular lattice, lies in the range $400-420$ nm. The two holes limiting each cavity have a reduced radius $r_1= r_0-0.06a$ and are displaced away by $s=0.16a$, in order to increase the Q-factor \cite{akahane2003}. Also holes around the cavities are modified with a period $2a$ by $r_2= r_0+0.05 a$ in order to improve the beaming quality of the emitted light and hence the collection efficiency \cite{PhysRevB.79.041101}; consequently, beaming holes inside the barrier have radius $r_4= r_2 (1+h)$. Only samples with $h\leq 0$ have been realized, which suffices to largely tune the inter-cavity coupling strength $g$. The resonance wavelengths of the samples range from $1500$ nm to $1600$ nm depending on the lattice period. The quality factor of the cavity resonances is $Q\sim 4000$ at $\lambda=1580$ nm, i.e. far form the QW absorption.  

Two kind of experimental characterizations have been carried out: reflectivity spectra and photoluminescence experiments, both with controlled spatial positions of the illumination spots. For the reflectivity spectra a single mode tunable laser is used, the reflected signal is coupled into a single mode optical fiber and sent to a femtowatt photodetector; the background reflectivity is highly suppressed using polarization optics (see Methods). In these experiments the cavity wavelengths, lying in the range $1560-1600$ nm, are red-detuned from the QW absorption, and the illumination power is low enough to be considered as linear reflectivity experiments. Hence, these can be interpreted as optical characterizations of linear Hermitian modes [Fig. \ref{fig2}(d)]. Unlike standard resonant scattering experiments leading to Fano resonances, the reflectivity background suppression allows us to clearly identify modes as Lorentzian-like peaks. 

The spectral position of the measured resonances is in very good agreement with the FDTD calculations of Fig. \ref{fig2}(b). Noticiably, three modes are clearly distinguished for $h=0\%$ and $h=-5\%$, corresponding to the low sublattice detuning region, while the middle peak is not apparent for $-25\%\leq h \leq-10\%$. This interval is within a crossover region where $g$ is small and changes sign [$-25\%\leq h \leq-5\%$ corresponding to $-2.17 \, \mathrm{THz}\leq g \leq 1.26\, \mathrm{THz}$, dashed region in Fig. \ref{fig2}(b)], with a crossing point ($g=0$) at $h\approx -15\%$. As a consequence, $M1$ and $M2$ frequencies are slightly split in this region. Finally, $M2$ resonances re-emerge for $h\leq-30\%$. While this $M2$ mode is expected to remain a dark mode and possess chiral symmetry in the Hermitian limit despite the large sublattice detuning, it will no longer be protected by NHPH symmetry, as will be discussed in the following. 

%{\color{red} We will introduce here Fig 2c and d. Fig 2c will be the intensity map and Fig 2d the the reflectivity as a function of h}
%%Figure 2(d) shows a spectral intensity map under spatial scanning of the $x$ position of the sample... DISCUSSION HERE

\section{Direct observation of the non-Hermitian zero-mode}
\label{S3}

\begin{figure}[!h]
\centering
\includegraphics[trim=6.5cm 2cm 0.5cm 2cm,clip=true,scale=0.55,angle=0,origin=c]{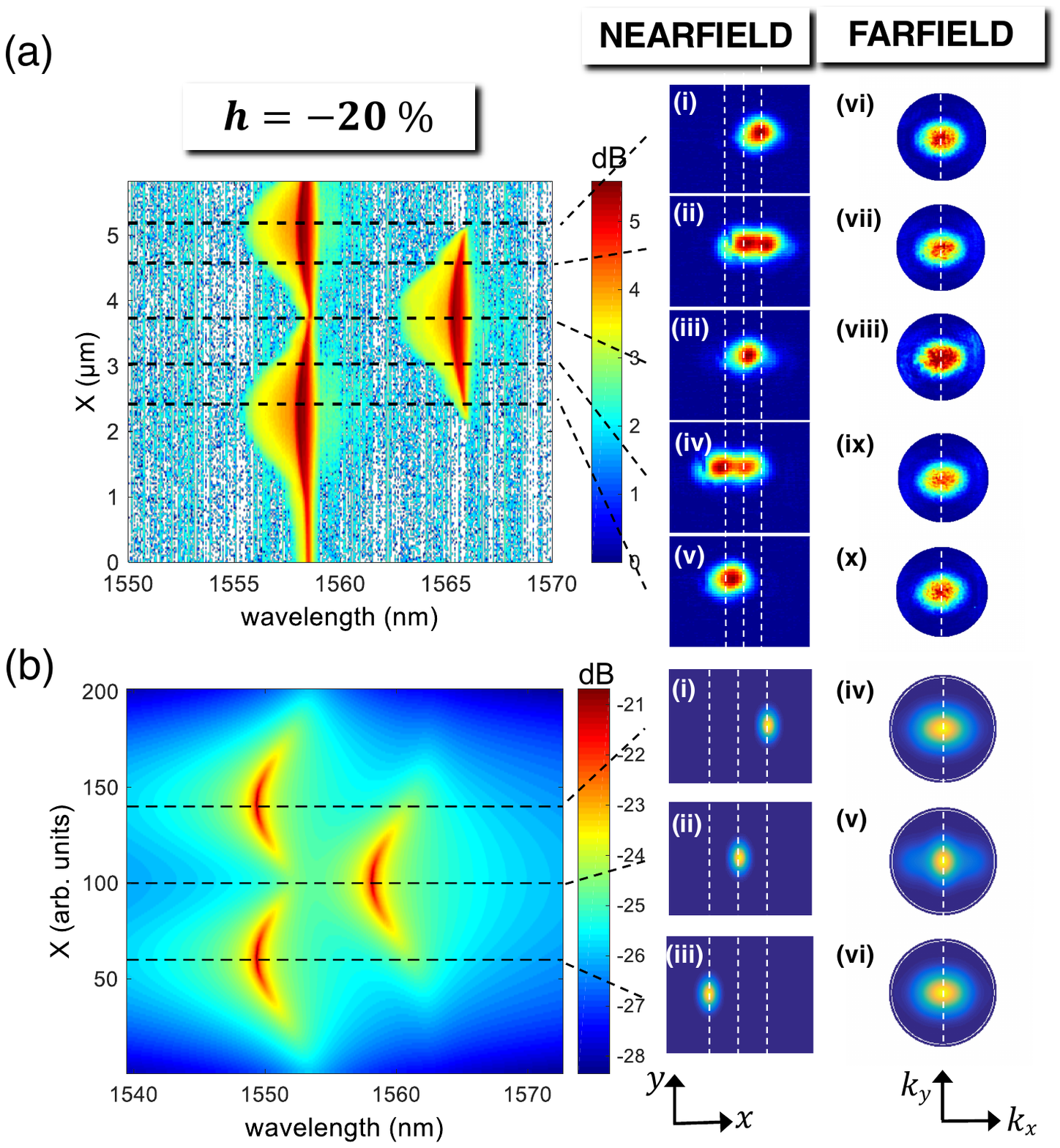}
\caption{
Spatially-resolved photoluminescence measurements in the large detuning regime. (a) Experimental results showing spectral intensity maps upon spatial scanning of a pump spot for $a = 416$ nm and $h=-20\%$, and (b) nonlinear CMT predictions. The horizontal position of the sample is changed by means of the piezo-electric voltage. Nearfield [(a)-i to (a)-v, (b)-i to (b)-iii] and farfield [(a)-vi to (a)-x, (b)-iv to (a)-vi] images are displayed at the selected pump spot positions marked with horizontal dashed lines. 
%(g) and (h) farfield and nearfield image of the lowest energy mode only respectively for Vhorz = 9.8 V, (c) and (d) farfield and nearfield image of the middle mode only for $V_{horz} = 14$ V , (i) and (j) farfield and nearfield image of the middle mode only for $V_{horz} = 5.7$ V. For the sample $a= 412$ nm and  $h  = -30 \%$:  (k), (m), (o), (q), (s) farfield images for all visible modes at pump position $V_{horz} = 6.2$ V, $V_{horz} = 9$ V, $V_{horz} = 11.8$ V, $V_{horz} = 14.3$ V, $V_{horz} = 16$ V respectively. (l), (n), (p), (r), (t) nearfield images for all visible modes at pump position $V_{horz} = 6.2$ V, $V_{horz} = 9$ V, $V_{horz} = 11.8$ V, $V_{horz} = 14.3$ V, $V_{horz} = 16$ V
}  
\label{fig3}
\end{figure}

\begin{figure}[!h]
\centering
\includegraphics[trim=6.5cm 2cm 0.5cm 2cm,clip=true,scale=0.55,angle=0,origin=c]{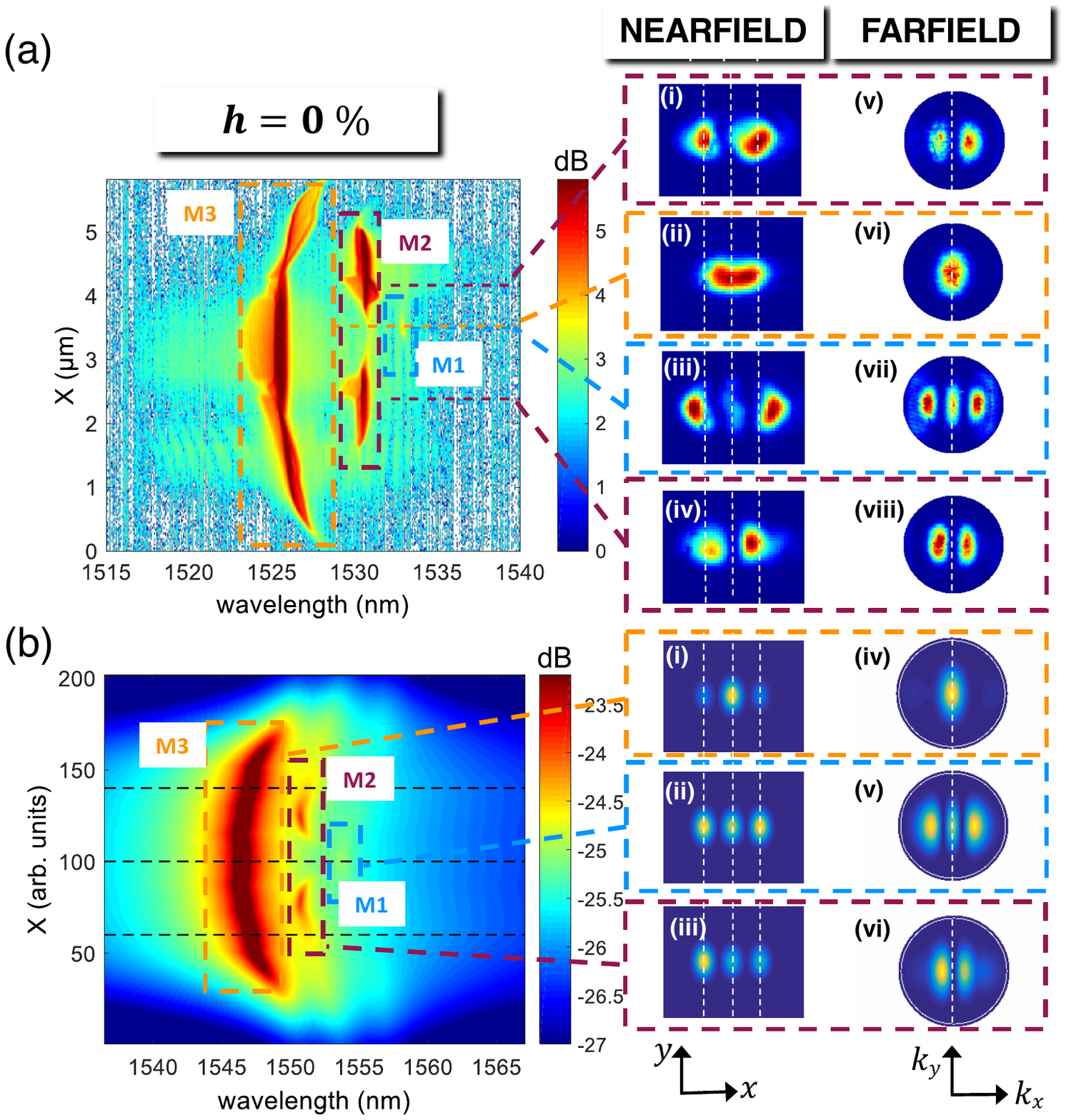}
\caption{
Observation of the zero mode in the low detuning regime. (a) Experimental results showing spectral intensity maps upon spatial scanning of a pump spot for $a = 408$ and $h=0\%$, and (b) nonlinear CMT predictions. The horizontal position of the sample is changed by means of the piezo-electric voltage.  
Nearfield [(a)-i to (a)-iv, (b)-i to (a)-iii] and farfield [(a)-v to (a)-viii, (b)-iv to (a)-vi] images are displayed at the selected pump spot positions marked with horizontal dashed lines. Spectral filters are used in order to remove contributions from other modes, the spectral bandwidth being represented by the horizontal extension of dashed boxes. 
%(g) and (h) farfield and nearfield image of the lowest energy mode only respectively for Vhorz = 9.8 V, (c) and (d) farfield and nearfield image of the middle mode only for $V_{horz} = 14$ V , (i) and (j) farfield and nearfield image of the middle mode only for $V_{horz} = 5.7$ V. For the sample $a= 412$ nm and  $h  = -30 \%$:  (k), (m), (o), (q), (s) farfield images for all visible modes at pump position $V_{horz} = 6.2$ V, $V_{horz} = 9$ V, $V_{horz} = 11.8$ V, $V_{horz} = 14.3$ V, $V_{horz} = 16$ V respectively. (l), (n), (p), (r), (t) nearfield images for all visible modes at pump position $V_{horz} = 6.2$ V, $V_{horz} = 9$ V, $V_{horz} = 11.8$ V, $V_{horz} = 14.3$ V, $V_{horz} = 16$ V
}  
\label{fig4}
\end{figure}

In addition to the reflectivity spectra of the previous Section, which characterize Hermitian modes, we have also performed photoluminescence (PL) experiments where the pump laser wavelength is now $\lambda=980$ nm: the laser beam is mainly absorbed in the quantum barriers and thus it can be considered as an incoherent pump. We use a pulsed laser ($100$ ps-duration and $1$ MHz repetition rate) in order to reduce thermal effects. As in the reflectivity experiments, the pump spot is focused down to nearly the diffraction limit so as to achieve a pump configuration with a localized gain profile across the cavity array, meaning that essentially one cavity is pumped when aligning the pump beam at its center. The radiated PL is collected in the free space and spectrally resolved with a spectrometer coupled to an InGaAs 1D detector array. A piezoelectric-driven stage holding the sample allows us to externally control the sample position with respect to the pump spot with sub-micron resolution. The results are shown in Figs. \ref{fig3}-\ref{fig4}.

The spectral intensity as a function of the sample position reveals two distinct typical patterns depending on the value of  the barrier parameter $h$. 
For $h<  -5\%$,  the detuning $|\Delta \omega(h)|$ is larger that the coupling strength $|g(h)|$; we can then expect that the two extreme cavities become effectively decoupled from the central one, specially for $h\leq -20\%$. The spectral map of Fig. \ref{fig3}(a) ($h=-20\%$)  is consistent with this prediction: as long as the pump excites the QWs in one of the extreme cavities, a resonant mode is observed at $\lambda_0\approx1558.2$ nm, and when pumping the middle cavity a mode red shifted by $\Delta \lambda \approx 7.7$ nm$\sim 6$ THz comes out, consistent with the $|\Delta \omega(h=-20\%)|\approx 7$ THz detuning obtained from the numerical simulations. 
%In order to validate this analysis we have measured far-field radiation patterns together with near-field imaging [Fig. \ref{fig3}(a)-vi to (a)-x]. 
The near-field images [Fig. \ref{fig3}(a)-i to (a)-v] show that the emission essentially comes from the pumped cavity provided only one mode is excited. Note that two near-field lobes can be observed for intermediate positions, where two resonances are simultaneously present in the spectrum [Fig. \ref{fig3}(a)-ii,iv]. The far-field images [Fig. \ref{fig3}(a)-vi to (a)-x] confirm this observation always revealing only one central lobe, consistent with a localized cavity mode. It is even the case for two near-field lobes, for which the radiation pattern becomes the incoherent superposition of two centered single-cavity far-field lobes [Fig. \ref{fig3}(a)-vii,ix]. 

This analysis allows us to conclude that the two external cavities are decoupled from the central one for large sublattice detuning. The experimental results are in very good agreement with CMT calculations including carrier-induced refractive index changes, that blue-shift the cavity resonance as a function of the pump power. We will refer to such a more realistic model for a semiconductor cavity as {\em nonlinear} CMT in the sense of a carrier-induced Kerr effect by an incoherent pump beam [Fig. \ref{fig3}(b), see Section IV of the Supplementary Material for further details].

Interestingly, although the zero mode would still exist in the case of large $\Delta \omega$ in Hermitian systems, it is no longer observable in our non-Hermitian system with detuning using a single localized pump spot. This is because Hermitian zero-modes warranted by sublattice symmetry are dark ones; therefore, they remain unaffected if the detuning takes place only in the cavities where the amplitude of the zero-modes is zero. On the other hand, the non-Hermitian zero modes are not dark ones in general. Here in our non-Hermitian system, 
%such a dark zero-mode would have experienced unequal pump power in its bright cavities (i.e., the top and bottom cavities) 
a single localized pump spot results in an imaginary detuning that acts together with the real (frequency) detuning to eliminate the NHPH symmetry and its zero-modes (see Supplementary material, Section I). Therefore, within this high sublattice detuning regime the system is generally not protected neither by sublattice nor by particle-hole symmetry and as a consequence there is no zero-mode that can be exploited.

\begin{figure*}[!h]
\centering
\includegraphics[trim=0cm 2cm 0cm 1cm,clip=true,scale=0.6,angle=0,origin=c]{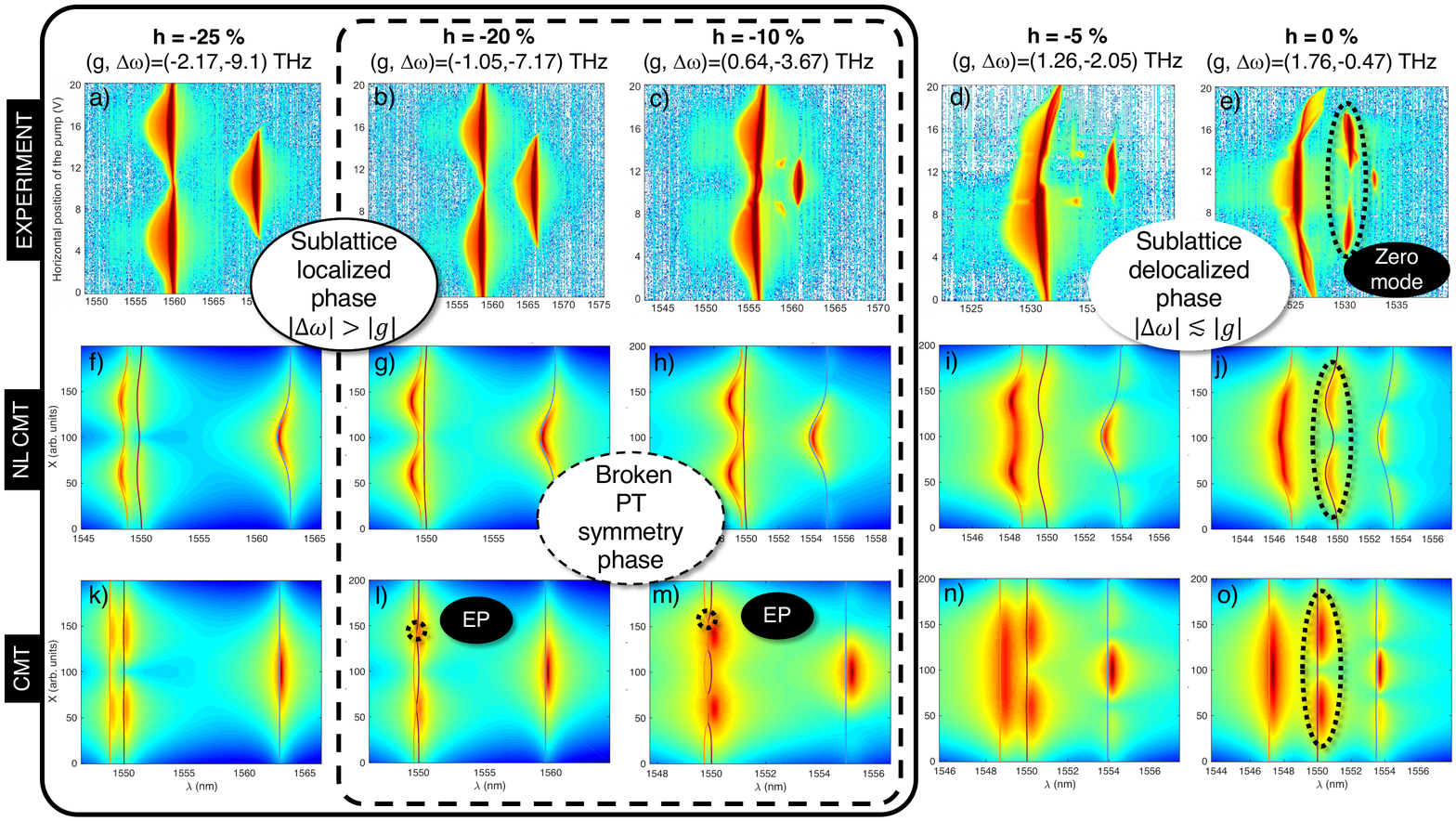}
\caption{
Phase diagram underlying the transition from sublattice delocalization and zero modes, to sublattice localization and mode coalescence. From (a) to (e): experimental PL intensity maps under pump spot position scanning across the coupled cavity system. (a)-(c) $a=416$ nm, $P_{pump}=0.8\,\mu$W; (d)-(e) $a=408$ nm, $P_{pump}=1.1\,\mu$W. From (f) to (j) nonlinear, and from (k) to (o) linear CMT calculations, with parameters $Q=4000$, $\lambda_0=1550$ nm, $\alpha_H=3$, $\sigma=30$. (f) and (k) $\gamma=1.05$, (g)-(h) and (l)-(m) $\gamma=1$, (i) and (n) $\gamma=1.2$, (j) $\gamma=1.45$ and (o) $\gamma=1.55$. 
}  
\label{fig5}
\end{figure*}

To experimentally address a zero-mode, we change the barrier parameter so that to enter into the low detuning regime, $h\sim 0\%$. 
Within this regime the spectral intensity pattern totally differs from the large detuning case where localized modes prevailed [Fig \ref{fig3}(a)]. 
%As the spatially selective pump 
%gets closer to one of the cavity, the wavelengths corresponding to the localized mode of said cavity will be shifted towards higher energies. This phenomenon can also be observed when pumping single cavities: when pumping below threshold the cavity frequency will be shifted towards the blue. This blue-shift is due to the properties of active QWs : in Eq(4) it is the result of the  non-linear frequency shift from transparency 〖Δω〗_NL (N).
For $-5\% \lesssim h\leq 10\%$ , $|\Delta \omega (h)| \lesssim |g(h)|$ [Fig \ref{fig2}(b)], and the three coupled cavities effectively behave as a whole. In the spectral intensity map [Fig. \ref{fig4}(a), $h=0\%$] 
%{\color{blue}Comment: Here I would suggest pointing out explicitly, with the explanation commented out above in the tex file,  that the intensities of these three modes are different from their linear counterparts.}) 
we can observe a pattern with the mode of highest energy $M3$ being excited independently of the sample position; it has a higher intensity compared to the two other modes. The central mode $M2$, on the other hand, attains two maxima in between the extreme cavities and the central one. The lowest energy mode $M1$ is the weakest one and it is only observed when pumping near the central cavity, a feature that was already present in the simplified calculation of Fig. \ref{fig1}(c).  
%For values of h in between -20 \% and -5 \% the experimental spectral analysis of the modes of the 3-coupled cavity lattice is inconclusive, as it is highly dependant on the lattice period and might even feature more modes than the expected three. We consider it a transitional zone in between high and low detuning regimes. 

In these conditions we have measured the near-field and far-field patterns setting the pump spot positions at the local maxima of the modes and using pass-band filters to filter out all other spectral components. From the near and far-field images of the highest [$M3$, Fig. \ref{fig4}(a)-ii and (a)-vi] and lowest [$M1$, Fig. \ref{fig4}(a)-iii and (a)-vii] energy modes we conclude that those are approximately symmetric modes, with $M1$ being the fundamental one, in agreement with the CMT and FDTD calculations. In particular, the nearfield images  [Figs. \ref{fig4}(a)-ii and (a)-iii] show that the emission of those two modes comes from all three cavities, with a higher intensity in the middle for $M3$ and in the extreme ones for $M1$. The central mode, on the other hand, features a far-field intensity node at the center [$M2$, Fig. \ref{fig4}(a)-v,viii]; its near-field is more intense in the two extreme cavities, [Fig. \ref{fig4}(a)-i,iv], while the intensity is below our detection limit in the central cavity region. These observations are compatible with the non-Hermitian zero mode $M2$ of Section 1. In particular, the $\pi/2$ phase jump between adjacent cavities predicted for a non-Hermitian zero mode is translated into a $\pi$ phase difference between the two extreme cavities, giving rise to an antisymmetric-like far-field profile, as it has already been observed for two coupled cavities \cite{Haddadi_2013}. 

Figure \ref{fig4}(b) shows the nonlinear CMT calculations. The carrier-induced refractive index effects have an important impact on the PL map, as compared to Fig. \ref{fig1}(c). Although the main qualitative features are already captured by a linear non Hermitian CMT, in the nonlinear CMT the blue-detuned mode $M3$ strongly dominates over the two other modes. This can be explained as a consequence of the frequency blue-shift of a cavity resonance under optical pumping: we can therefore expect that the blue most detuned hybrid mode will be more efficiently excited, since its spectral overlap with the pumped cavity resonance increases. This explains the enhancement of the high energy mode $M3$ in this low sublattice detuning regime, even though the zero mode is clearly measurable. Importantly, the zero mode might be brought to laser operation as long as a two spot pumping scheme is implemented, as will be reported elsewhere.

\section{Zero mode coalescence and phase transitions}
\label{S4}

In this section we unveil the underlying physical mechanisms that lead to the extinction of zero modes as the barrier parameter $h$ is decreased from $h=0\%$ to $h=-25\%$, i.e. as the intercavity coupling $|g|$ goes from above to below the sublattice detuning $|\Delta\omega|$. 
Indeed, the non Hermitian zero mode is only observed in the $|\Delta\omega|\lesssim |g|$ regime (Fig. \ref{fig4}), while it is missing in the $|\Delta\omega|> |g|$ regime (Fig. \ref{fig3}). First of all, let us recall that, in the large sublattice detuning regime, NHPH symmetry no longer warrants zero modes for single cavity pumping. But even if $M2$ looses its symmetry protection, the question arises of whether $M2$ still exists but remains undetectable, or it coalesces through a phase transition. Let us also recall that, in the Hermitian limit, the zero mode is observable for large sublattice detuning [Fig. \ref{fig2}(d), $h=-30\%$, and $h=-35\%$].

The full picture of the different non Hermitian phases as a function of the barrier parameter is represented in Fig. \ref{fig5}; this completes the experimental PL maps for various $h$ parameters [Fig. \ref{fig5}(a)-(e)], and depicts nonlinear [Fig. \ref{fig5}(f)-(j)] as well as linear [Fig. \ref{fig5}(k)-(o)] CMT calculations. The experimental cases already analyzed in the previous section are reproduced in Fig. \ref{fig5}(b) ($h=-20\%$) and in Fig. \ref{fig5}(e) ($h=0\%$). We first identify two phases, corresponding to effective coupling/decoupling of the central cavity with respect to the two extreme ones: the {\em sublattice delocalized} phase corresponds to $|\Delta\omega|\lesssim |g|$ ($-5\%\lesssim h \leq 0\%$), in which $M2$ is observable and becomes the zero mode because of NHPH symmetry; conversely, the {\em sublattice localized} phase corresponds to $|\Delta\omega|> |g|$ ($h \leq -10\%$), in which $M2$ is no longer observable. Interestingly, within this sublattice localized phase, there is a sub-region corresponding to the crossover of $g$, from positive to negative, vanishing at $h\approx -15\%$. At such a crosssover $M1$ and $M3$ exchange symmetries, in the sense that $M1$ goes from a quasi symmetric mode with zero phase jumps between the cavities for $g>0$, to a quasi symmetric mode with $\pi$ phase flips between the cavities for $g<0$. 

More importantly, for very small $g$ a parity time symmetry breaking is predicted as one extreme cavity is pumped. Within this {\em broken PT symmetry} phase ($-20\%\lesssim h \lesssim -10\%$), the central cavity (sublattice $B$) --already effectively decoupled from the two extreme ones-- does not play an important role; at the same time, the two extreme ones (sublattice $A$) weakly interact with each other, such that a gain unbalance may undergo an exceptional point (EP). This is depicted in Figs. \ref{fig5}(l) and (m): the real part of the eigenvalues of modes $M2$ and $M3$ undergo EPs when pumping in the proximity of an extreme cavity. Within a pumped cavity of sublattice $A$, as it is apparent in Fig. \ref{fig5}(m), the real parts of the eigenvalues of $M2$ and $M3$ coalesce in a single branch; there, only one mode is observable --the one with higher gain--, and it is localized in the pumped cavity. 

Of course, in a semiconductor material under gain/loss operation, parameters can only be tuned to some extent near an EP.  This is because, in addition to imperfect symmetries in real systems, semiconductors exhibit carrier-induced frequency shift that, in general, explicitly breaks the inversion symmetry of the real part of the dielectric constant. Therefore, the otherwise EP bifurcation results in an imperfect symmetry breaking [Figs. \ref{fig5}(g) and (h)]~\cite{garbin_asymmetric_2021}. As already discussed, such a frequency shift enhances light localization in an extreme cavity when it is optically pumped. Therefore, within this broken PT symmetry region, light localization in the real device is a combination of both the underlying PT symmetry breaking and nonlinear effects, while outside this region [Figs. \ref{fig5}(a), (f) and (k)], localization in one extreme cavity arises because of pure carrier induced refractive index effects. As a matter of fact, even if not detected in Figs. \ref{fig5}(a), the theory still predicts the $M2$ mode to be observable [see Figs. \ref{fig5}(f) and (k)]. 

\section{Conclusion}
\label{S5}

We have reported on the direct observation of non Hermitian zero modes warranted by non Hermitian particle hole (NHPH) symmetry in a minimal cavity array: three coupled photonic crystal nanocavities in a gain/loss configuration under spatially localized optical pumping. Because the number of cavities is odd, the Lieb's theorem ensures the existence of a zero mode, and there is no need to generate them through, for instance, NHPH symmetry restoration. The nature and properties of these non Hermitian zero modes differ from their chiral Hermitian counterparts in two main aspects. Firstly, unlike Hermitian zero modes, non Hermitian ones are more robust in the sense that they are not restricted to the origin of the complex plane, but they may exist along the imaginary axis, still benefiting from symmetry protection. More specifically, and analogously to chiral modes, they are immune to random coupling perturbations in the cavity array. Secondly, although non Hermitian zero modes are not dark ones, in the sense that there is no light extinction in one of the sublattices, these zero modes feature $\pi/2$ phase jumps between adjacent cavities. These constitute unambiguous physical signatures that distinguish their wavefunctions from any other ones in the array, and enables experimental protocols to detect them. Here we have shown that a photoluminescence (PL) intensity map under spatial scanning of the pump spot provides a clear fingerprint of zero modes in the form of PL maxima in between cavities. At those PL maxima, a Fourier imaging technique allows us to identify phase jumps as destructive interference in the far field. Specifically, the $\pi/2$ phase jumps between contiguous cavities result in a $\pi$ phase difference between the extreme cavities, which can be detected as nodes at the far field center ($k=0$). 

We have identified different regimes that arise as the coupling barrier of the photonic molecule is systematically modified through the perturbation parameter $h$, that varies the hole radius of a row within the photonic barrier. Such a barrier modification simultaneously changes the coupling $g$ and the sublattice detuning $\Delta \omega$, giving rise to two important regimes: $|\Delta \omega|> |g|$ leading to sublaticce localization, and $|\Delta \omega|\lesssim |g|$ leading to sublattice delocalization and zero modes. Remarkably, we have identified the transition from zero mode observation to its extinction as $h$ is decreased as a combination of both an underlying parity time symmetry breaking and carrier induced blue shift nonlinearities, that strongly localize the PL in a single cavity. Outside this PT-symmetry broken phase, and within the strong sublattice detuning region, the Hermitian zero modes do exist even though the system is no longer chiral, which we have shown from reflectivity resonances in a linear resonant scattering experiment ($h\leq -35\%$). On the other hand, the non Hermitian zero mode cannot be observed in this regime with a single pump spot because of symmetry mismatch under symmetric excitation. Moreover, carrirer induced blue shift effects enhance the PL emission within the highest energy mode, hindering the central resonance $M2$. 

In spite of the tremendous theoretical advances in the non-Hermitian domain, observations of photonic zero modes have been restricted so far to waveguide or cavity arrays that ressemble their Hermitian counterparts, such as the Su-Schrieffer-Heeger (SSH) lattice. Although experimental demonstrations of non Hermitian phenomena such as PT symmetry breaking and exceptional points are coming to maturity, with special focus in two resonator systems, the physical realization of non Hermitian symmetries beyond PT such as NHPH in large cavity arrays --a unique playground for non Hermitian photonics-- is still in its infancy. We believe that further developments in this direction would enable promising applications of non Hermitian symmetry protected modes, ranging from laser array mode engineering to photonic computing.  

\section*{Acknowledgments}
This work is supported by a public grant overseen by the French National Research Agency (ANR) as part of the "Investissements d'Avenir" program (Labex NanoSaclay, reference: ANR-10-LABX-0035) and by the ANR UNIQ DS078. B. G. and A. M. Y. would like to acknowledge the financial support from the European Union in the form of Marie Sk\l odowska-Curie Action grant MSCA-841351. L.G. acknowledges support by National Science Foundation under grant No. PHY-1847240.

\section*{Methods}

\subsection*{Linear reflectivity experiments}

We characterize the behavior of the linear Hermitian modes in our system by performing linear reflectivity experiments as briefly described in the main text. The quasi-resonant injection beam is obtained from a Tunics T100s-HP, and its polarization is managed by means of a polarizing beam splitter combined to a subsequent half-wave plate that ensures linear horizontal polarization. The injection beam reaches the injection/detection beam splitter (80~\% reflectivity), a half-wave plate and is finally injected through a 0.95 numerical aperture microscope objective (Olympus MPLAN x100 IR). The fast axis of the half-wave plate is rotated at 22.5$^\circ$ from the injection beam allowing an injection at 45$^\circ$ from the cavities polarization and an according polarization separation between the non-injected and the injected beams. An additional lens with 100~cm focal is located prior to injection so as to facilitate mode matching with the cavities. We then separate the non-injected and injected beams after the injection/detection beam splitter using a half-wave plate---in order to select the detected polarization---and a polarizing beam splitter. We finally inject a monomode optical fiber with the three cavity systems' emission and detect the signal using a Femtowatt Photoreceivers (New Focus 2153) connected to a 12-bits oscilloscope (Tektronix MSO64).

During a single realization of the experiment, the injection beam is first centered onto the considered nanocavity array, and subsequently sweep over either 20 or 40~nm with different starting wavelengths [as presented in \figurename~{\ref{fig2}} (d)] depending on the barrier parameter.

%\bibliography{sample}
\bibliography{switch_Ale}

\end{document}